\documentclass[showpacs,amsmath,amssymb,superscriptaddress,twocolumn,prd,aps,10pt]{revtex4-1}
\usepackage{graphicx,amsmath,color}
\usepackage{graphicx}
\usepackage{psfrag}
\input{tcilatex}
\newcommand{\ve}{\varepsilon}
\newcommand{\vp}{\varphi}

\begin{document}
\title{A topologically flat thick 2-brane on higher dimensional black hole backgrounds}
\author{Viktor G.~Czinner}
\email{czinner@rmki.kfki.hu}
\affiliation{Department of Mathematics and Applied Mathematics, University of Cape Town,
\\ Rondebosch, 7701, South Africa;}
\affiliation{Department of Theoretical Physics, MTA KFKI Research Institute for Particle 
and Nuclear Physics,\\ Budapest 114, P.O.~Box 49, H-1525, Hungary}

\begin{abstract}
We present a numerical solution for a topologically flat 2-dimensional thick brane on 
a higher dimensional, spherically symmetric black hole background. Present solution is the last, missing 
part of the complete set of solutions for the thickness corrected brane - black hole problem in 
arbitrary number of dimensions. We show that the 2-dimensional case is special compared to all the higher 
dimensional solutions in the topologically Minkowskian family as being non-analytic at the axis of the system. 
We provide the numerical solution in the near horizon region and make a comparison with the infinitely thin case.  

\end{abstract}
\pacs{04.70.Bw, 04.50.-h, 11.27.+d}

\maketitle

\section{Introduction}

The study of higher dimensional black holes, branes and their interactions is an active 
field of research in several different areas of modern theoretical physics 
\cite{flactan,Frolov,MMT1,Filpe}. One interesting direction, which has been first introduced 
by Frolov \cite{Frolov}, is to consider a brane - black hole (BBH) toy model for studying 
merger and topology changing transitions in higher dimensional classical general relativity 
\cite{Kol1,Kol2}, or in certain strongly coupled gauge theories \cite{MMT1,MMT2} 
through the AdS/CFT correspondence \cite{Maldacena}. Generalizations of the BBH model by 
studying thickness corrections to the Dirac-Nambu-Goto effective brane action 
\cite{Dirac,Nambu,Goto} from higher order curvature terms, have also been studied recently, 
first by perturbative approaches \cite{FG,CF}, and later within an exact description \cite{Cz}.

The results of the perturbative approaches concluded that there is a ''symmetry breaking'' 
between the two topologically different solution family, as regular perturbative solutions do not exist 
for Minkowskian embedding topologies except in the special case of a 2-brane. The problem, 
on the other hand, can be solved regularly for any brane dimensions in the black hole embeddings. 
This virtual ''symmetry breaking'' phenomenon obtained a simple resolution in \cite{CF}, where 
it was pointed out that perturbative thick solutions break down around their thin counterparts 
for Minkowski topologies, because the thin solutions are not analytic at the axis of the system. 
Motivated by this observation, in \cite{Cz}, a general family of thick solutions could be 
provided for both topologies within a non-perturbative numerical approach for all but one exceptional 
case. The exception, mysteriously, turned out to be the same case, where the regular 
perturbative solution existed, namely the 2-dimensional, topologically flat case.

The above findings of \cite{Cz} naturally raised the question: How can a regular perturbative solution exist
in the same single case where the regular non-perturbative solution can not be found? 
       
In the present paper we provide the answer to this question and obtain the so far missing  
solution of the topologically flat 2-brane in arbitrary number of bulk dimensions for the 
thick-BBH system. By including this solution, we complete the full set of solutions of the problem 
that we started presenting first with a perturbative approach in \cite{CF} and continued with a
non-perturbative description in \cite{Cz}. The present work, therefore, is the final part of the series of papers we 
addressed to the thickness corrected BBH problem and we kindly refer the reader also to 
\cite{CF,Cz} for the more detailed model setups and for all those  definitions, notations 
and results that might be missing here and would make the present paper completely self-contained. 
 
The plan of the paper is as follows. In Sec.~II, we provide a short overview on the 
thin- and thick-BBH model setups. In Sec.~III we obtain the special case of the 
2-brane equation, and in Sec.~IV we analyze its regularity conditions for flat topology. 
In Sec.~V we provide the non-perturbative, numerical solution of the problem in the near horizon region, 
and in Sec.~VI we draw our conclusions.

\section{The brane - black hole system}

Let us overview quickly the most important properties of the thin-BBH system introduced by Frolov 
in \cite{Frolov}, and its thickness corrected generalization provided in \cite{CF}. 

\subsection{The thin model}

We consider static brane configurations in the background of a static, spherically 
symmetric bulk black hole. The metric of an $N$-dimensional, spherically symmetric black hole spacetime is
\begin{equation}
ds^2=g_{ab}dx^adx^b=-fdt^2+f^{-1}dr^2+r^2d\Omega_{N-2}^2\ ,
\end{equation}
where $f=f(r)$ and $d\Omega_{N-2}^2$ is the metric of an $N-2$ dimensional unit sphere. One can define 
coordinates $\theta_i (i=1,\dots, N-2)$ on this sphere with the relation
\begin{equation}
d\Omega_{i+1}^2=d\theta_{i+1}^2+\sin^2\theta_{i+1} d\Omega_i^2 \ .
\end{equation}
The explicit form of $f$ is not important, it is only assumed that $f$ is zero at the horizon $r_0$, and it 
grows monotonically to $1$ at the spatial infinity $r\rightarrow \infty$, where it has the asymptotic form 
\cite{Tangherlini},
\begin{equation}\label{f}
f=1-\frac{r_0}{r^{N-3}}\ .
\end{equation}

In the zero thickness case the test brane configurations in an external gravitational field can be obtained 
by solving the equation of motion coming from the Dirac-Nambu-Goto action \cite{Dirac, Nambu, Goto},
\begin{equation}\label{action0}
S=\int d^D\zeta\sqrt{-\mbox{det}\gamma_{\mu\nu}}\ , 
\end{equation}
where $\gamma_{\mu\nu}$ is the induced metric on the brane
\begin{equation}
\gamma_{\mu\nu} =g_{ab}\frac{\partial x^a}{\partial \zeta^{\mu}}
\frac{\partial x^b}{\partial \zeta^{\nu}}\ ,
\end{equation}
and $\zeta^{\mu}(\mu=0,\dots ,D-1)$ are coordinates on the brane world sheet. The brane tension does not enter 
into the brane equations, thus for simplicity it can be put equal to $1$. It is also assumed that the brane is 
static and spherically symmetric, and its surface is chosen to obey the equations
\begin{equation}
\theta_D=\dots =\theta_{N-2}=\pi/2\ .
\end{equation}
With the above symmetry properties the brane world sheet can be defined by the function $\theta_{D-1}=\theta(r)$ 
and we shall use coordinates $\zeta^{\mu}$ on the brane as
\begin{equation}
\zeta^{\mu}=\{t,r,\phi_1,\dots,\phi_{n}\}\quad \mbox{with} \quad n=D-2 \ . 
\end{equation}
The parameter $n$ denotes the number of dimensions in which the brane
is rotationally symmetric. In this paper we consider the special case
of $n=1$, i.e.~a 3-dimensional brane world sheet, that is a 2-dimensional,
axisymmetric brane embedded into the higher dimensional black hole
spacetime.

With this parametrization the induced metric on the brane is
\begin{equation}
\gamma_{\mu\nu} d\zeta^{\mu}d\zeta^{\nu}=-fdt^2+\left[\frac{1}{f}+r^2{\dot\theta}^2\right]dr^2+r^2\sin^2\theta d\Omega_n^2,
\end{equation}
where, and throughout this paper, a dot denotes the derivative with respect to $r$, and the action (\ref{action0}) reduces 
to
\begin{eqnarray}
S&=&\Delta t \mathcal{A}_n\int\mathcal{L}_0\ dr\ ,\\
\mathcal{L}_0&=&r^n\sin^n\theta\sqrt{1+fr^2{\dot\theta}^2}\ ,\label{L0} 
\end{eqnarray}
where $\Delta t$ is the interval of time and $\mathcal{A}_n=2\pi^{n/2}/\Gamma(n/2)$ is the surface area of a unit $n$-dimensional 
sphere.

\subsection{Thickness corrections}
In the case of a thick brane, the curvature corrected effective brane action is obtained by Carter and Gregory 
in \cite{Carter}, and  the corrections to the thin DNG action are induced by small thickness 
perturbations as
\begin{equation}\label{action1}
S=\int d^D\zeta\sqrt{-\mbox{det}\gamma_{\mu\nu}}\left[-\tfrac{8\mu^2}
{3 \ell}(1+C_1R+C_2K^2)\right],
\end{equation}
where $R$ is the Ricci scalar, $K$ is the extrinsic curvature scalar of the brane and the coefficients $C_1$ and 
$C_2$ are expressed by the wall thickness parameter $\ell$ as
\begin{equation}
C_1=\frac{\pi^2-6}{24}\ell^2\ , \qquad C_2= -\frac{1}{3}\ell^2 .
\end{equation}
The parameter $\mu$ is related to the thickness by 
\begin{equation}
\ell=\frac{1}{\mu\sqrt{2\lambda}}
\end{equation}
which originates from a field theoretical domain-wall model where $\mu$ is the mass parameter and $\lambda$ is the 
coupling constant of the scalar field.

After integrating out the spherical symmetric part and the time dependence on the introduced static, spherically 
symmetric, higher dimensional black hole background, one obtains 
(see also \cite{CF})
\begin{eqnarray}\label{S}
S&=&\Delta t \mathcal{A}_n\int\mathcal{L}\ dr\ ,\\
\mathcal{L}&=&-\frac{8\mu^2}{3\ell}\mathcal{L}_0[1+\ve\delta]\ ,\label{L} 
\end{eqnarray}
where we introduced the notations 
\begin{eqnarray}\label{delta}
\ve=\frac{\ell^2}{L^2},\qquad \delta&=&aK^2+bQ\ ,
\end{eqnarray}
with
\begin{equation}\label{ab}
Q= K^a_bK^b_a,\quad a=\frac{\pi^2-14}{24}L^2,\quad b=\frac{6-\pi^2}{24}L^2\ .
\end{equation}
Here $L$ is the relevant dynamical length scale of the system which has to be large compared to 
the thickness parameter $\ell$ in order to (\ref{action1}) remain valid. The explicit expressions 
of the curvature scalars $K$ and $Q$ are given in (35) and (36) of \cite{CF}.

For a detailed introduction of both the thin- and thick-BBH systems, please refer to \cite{Frolov,CF}. 

\section{The 2-brane equation}

From this section on we will focus only on the case of the topologically flat or Minkowskian 2-brane.
In two dimensions ''topologically flat'' is synonymous with ''topologically Minkowskian'', and we 
retain both terms for the sake of elegant variation.

In order to obtain the 2-brane equation of motion, first we observe that the thickness corrected 
DNG-brane action is a function of the second derivative of $\theta$ and thus the Euler-Lagrange 
equation of the problem has the form (see for example \cite{Whittaker})
\begin{equation}\label{mel}
\frac{d^2}{dr^2}\left(\frac{\partial \mathcal{L}}{\partial \ddot\theta}\right)
-\frac{d}{dr}\left(\frac{\partial \mathcal{L}}{\partial \dot\theta}\right)
+\frac{\partial \mathcal{L}}{\partial \theta}=0\ .
\end{equation}
From (\ref{mel}) the actual equation of motion becomes
\begin{eqnarray}\label{fulleq}
\theta^{(4)}&+&T_1\theta^{(3)}
+T_2(\ddot\theta,\dot\theta,\theta,f^{(3)},\ddot f,\dot f,f,r)=0\ ,
\end{eqnarray}
where, in the 2-dimensional ($n=1$) case 
\begin{eqnarray}\label{T1}
T_1&=&\frac{1}{rfF^2}\left[\right.
6f+4r\dot f+2rf\cot\theta\dot\theta
-r^2f\left(4f+r\dot f\right)\dot\theta^2\nonumber\\
&+&2r^3f^2\cot\theta\dot\theta^3-10r^3f^2\dot\theta\ddot\theta \left.\right]\ ,
\end{eqnarray}
with
\begin{equation}
F=\sqrt{1+fr^2\dot\theta},
\end{equation}
and $T_2$ is given in the Appendix. 

As it is immediate to see, (\ref{fulleq}) is a $4th$-order, highly nonlinear equation,
and it is probably impossible to present its solutions in closed, analytic form. Hence,  
the goal of this paper, is to provide a regular, numerical solution of (\ref{fulleq}) 
in arbitrary bulk dimensions for flat topology. 

\section{Regularity and boundary conditions}

It was pointed out first in \cite{CF}, that the brane solutions obtained by Frolov in \cite{Frolov}
are regular but not analytic (or smooth) at the axis of the thin-BBH system for the Minkowski embedding branch. 
In fact they are not even differentiable at that point and thus belong to the class of $C^0$ 
functions only. According to this property, and since we found that the perturbative approach broke down around 
these solutions for the thick case,  we concluded that the thick brane solutions must behave significantly 
differently at the axis of the system, namely we expected them to be smooth there. It was surprising however that in the 
single case of the 2-brane, a regular perturbative solution existed. We gave a detailed analysis of this 
solution in \cite{CF}. 

In \cite{Cz}, approaching the problem by a new, non-perturbative numerical method, we looked for the missing
solutions of the Minkowski branch in the class of analytic functions. We obtained regular boundary conditions 
at the axis of the system by considering the series expansion of the exact $4th$-order equation of motion 
around $\theta=0$. With this method we successfully provided all the missing, topologically Minkowskian
solutions of the thick-BBH system, except in the curious case of the 2-brane again.  

It was obvious, of course, that a perturbative solution can not exist without the existence of a non-perturbative solution, 
and also since explicitly constructed, field theoretical domain wall solutions \cite{Fl1,Fl2} clearly exist in the 
case of the 2-brane, we suspected that the lack of this solution must lie somewhere in the validity 
of the applied method. Nevertheless, we were so enthusiastic with providing the whole family of the missing solutions
in the $C^{\infty}$ class at the axis, that it didn't occur in our mind at the time, that the 2-dimensional case might be special 
in the topologically Minkowskian family as being the only one which is non-analytic at the axis.

The main result of the present paper is the observation, that the 2-dimensional solution of the thick-BBH system 
is in fact a special one in the topologically Minkowskian family as being $C^0$ (or as we will soon see maximum 
$C^1$) function at the axis of the system. All the other dimensional solutions are $C^{\infty}$. This also explains 
the fact why it was only the 2-dimensional case where a perturbative solution could exist around the thin solution. 

In the remaining of this section we analyze the asymptotic behavior of (\ref{fulleq}) near the axis of the system.   
We obtain necessary boundary conditions from regularity requirements and show that these conditions can always be
fulfilled in order to obtain a regular solution.

\subsection*{Asymptotic analysis}

We know from the above considerations that a regular solution of the problem must exist although analytic
solution could not be found at the axis of the system. Thus the point $r_1$ on the axis, where $\theta(r_1)=0$, must 
be a \textit{regular singular point} of the differential equation (\ref{fulleq}). Even though (\ref{fulleq}) is 
highly nonlinear, general results from the theory of \textit{local} analysis of linear differential equations can be applied, 
because we know from physical considerations that (\ref{fulleq}) should not develop any nontrivial singular points in 
its domain. 

Hence, if a solution is not analytic at a regular singular point (see e.g.~\cite{BO}), its singularity must be either a pole 
or an algebraic or logarithmic branch point, and there is always at least one solution of the form 
\begin{equation}\label{alfaeq}
\theta(r)=(r-r_1)^{\alpha}A(r)
\end{equation}
where $\alpha$ is a number called {\it indical exponent} and $A(r)$ is a function which is analytic at $r_1$
and has a convergent Taylor series. 

In the general case $\alpha$ can be any number that solves (\ref{fulleq}).  In our specific case however, the thin 
solution, around which a regular perturbative solution existed, had the asymptotic form near $r_1$ 
(see \cite{Frolov,CF})
\begin{equation}
\theta(r)=\eta\sqrt{r-r_1}+\dots ,
\end{equation}
and similarly, the corresponding perturbative solution (see \cite{CF}) near the same point had the asymptotic 
form 
\begin{equation}\label{fi}
\theta_{thick}\equiv\theta_{thin}+\ve\vp=(\eta+\ve\kappa)\sqrt{r-r_1}+\dots ,
\end{equation}
where $\vp$ is the perturbation function and $\eta$ and $\kappa$
are coefficient functions defined in \cite{CF}. Thus, in order to obtain
the non-perturbative thick solution, we also chose $\alpha$ to be $1/2$, 
in accordance with the perturbative results. This choice will also have the advantage 
of naturally fixing the free boundary condition in the next section for a unique numerical solution. 

With $\alpha=1/2$, from (\ref{alfaeq}) we obtain that the asymptotic form of $\theta(r)$
near the axis is
\begin{eqnarray}\label{asymp}
\theta(r)&=&A_1\sqrt{r-r_1}+A_2(r-r_1)^{\frac{3}{2}}\\
&+&A_3(r-r_1)^{\frac{5}{2}}+A_4(r-r_1)^{\frac{7}{2}}+\dots\ .\nonumber
\end{eqnarray}
Plugging this expression and its derivatives into (\ref{fulleq}), one can obtain
the following asymptotic behavior 
\begin{eqnarray}
\frac{c_5}{(r-r_1)^{\frac{5}{2}}}+\frac{c_3}{(r-r_1)^{\frac{3}{2}}}+\frac{c_1}{\sqrt{r-r_1}}+c_0+\dots=0,\quad
\end{eqnarray}
where the coefficient functions $c_i$ are polynomial expressions of $A_i$ with dependences 
\begin{eqnarray}
c_5&=&c_5(A_1,A_2)\nonumber\\
c_3&=&c_3(A_1,A_2,A_3)\nonumber\\
c_1&=&c_1(A_1,A_2,A_3,A_4)\ .\nonumber
\end{eqnarray}
In order to obtain a regular solution, we need to require for the $c_i$ coefficient functions to 
disappear at $r_1$. From the explicit forms of $c_i$, one can find that the coefficient $A_1$ can 
be chosen freely, and once it's fixed, the remaining coefficients $A_2$, $A_3$ and $A_4$ can be 
computed from the requirements that $c_i(r_1)=0$. Consequently, the solution is uniquely determined by the
parameter $r_1$ (i.e.~the minimal distance parameter), and the explicit values of the coefficients
$A_2$, $A_3$ and $A_4$ can be immediately obtained from the $c_i(r_1)=0$ equations. (This procedure 
is of course necessary before the numerical setup.) For the question of existence, the $c_i(r_1)=0$ 
equations are always soluble, as it turns out that these are linear equations for the 
$A_i$ coefficients. The explicit forms of the coefficients $A_2$, $A_3$ and $A_4$, as successive 
functions of $A_1$ and $r_1$ are given in the Appendix.

In conclusion we found that a regular 2-dimensional solution can always be given for the exact problem,
and a unique solution is completely determined by the regularity requirements once the coefficient $A_1$ 
is fixed. Since we are free to choose $A_1$, it can also put to be $0$ for example. In this case the 
asymptotic form (\ref{asymp}) starts with
\begin{equation}
\theta(r)=A_2(r-r_1)^{\frac{3}{2}} + \dots,
\end{equation}
which solution is a $C^1$ function, however in all other cases the solution of (\ref{fulleq}) is $C^0$
at $r_1$.

\section{Numerical solution near the horizon}

For illustrating the obtained results, we provide the numerical solution 
of the 2-dimensional flat problem. With the experiences we gained from the analysis of the 
thick-BBH system in \cite{CF,Cz}, it is not too difficult to obtain the numerical solution 
here, after the initial conditions for this specific case has been clarified. 

As we discussed earlier, we are free to choose the coefficient $A_1$ in the asymptotic solution
(\ref{asymp}). Nevertheless, in order to be completely consistent with our previous perturbative 
results in the flat 2-brane case, we make the choice  
\begin{equation}
A_1=\eta+\ve\kappa
\end{equation}
that was forced upon us by regularity requirements for the perturbations. Having fixed this freedom,
the remaining three conditions $A_2$, $A_3$ and $A_4$ are determined, as discussed in the previous section,
and the corresponding numerical solution is unique.

In obtaining the numerical solution we used Mathematica${}^\circledR{}$ \texttt{NDSolve} function.
The integration range went from $r_1$ until $1000$ to check the accordance with the corresponding
perturbative solution.

The configurations of the perturbative thick 2-brane solutions in the near horizon region have been 
presented in \cite{CF}. Since the effects of the nonlinearities are really small even in the gravitationally
strong, near horizon region, the overall global picture of the thickness corrected brane configurations
remain very similar to the perturbative case. To make however the small effects visible, we plot on 
Fig.~1 and Fig.~2 the difference function 
\[
 \Delta\theta(r)=\theta(r)-\theta_{DNG}(r) ,
\]
of the present thick- and the original thin solutions in 4 and 5 dimensions. $\Delta\theta(r)$ is the exact 
analog of the perturbation function $\ve\vp(r)$ defined in (\ref{fi}).  
\begin{figure}[!ht]
\noindent\hfil\includegraphics[scale=1]{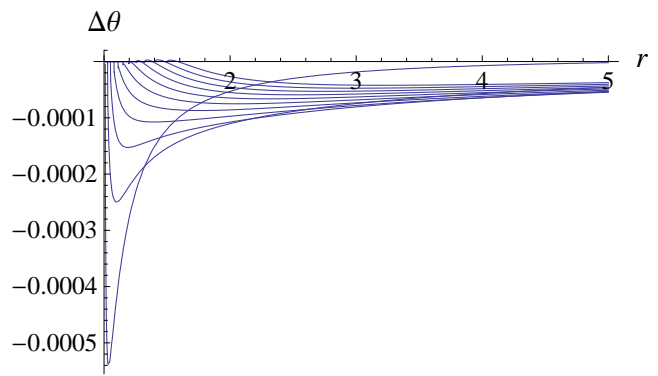} 
\caption{The picture shows a sequence of near horizon $\Delta\theta(r)$ curves in 4 dimensions with minimum 
horizon distance range $1.01\le r_1\le 2$.}
\end{figure}
\begin{figure}[!ht]
\noindent\hfil\includegraphics[scale=1]{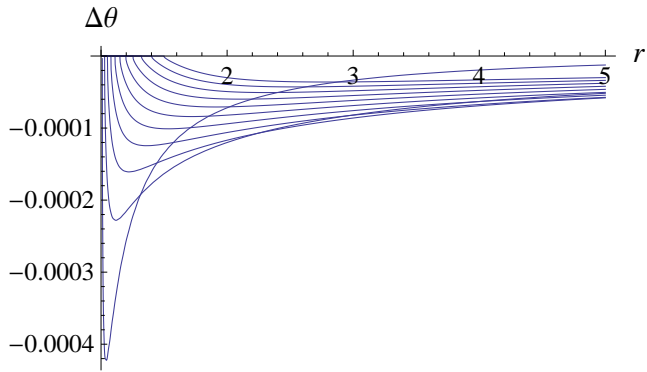} 
\caption{The same sequence of $\Delta\theta(r)$ curves as on FIG.1 in 5 dimensions.}
\end{figure}

Comparing the results with the corresponding plots of the perturbative solutions in \cite{CF}, we find that
the general behavior of the $\Delta\theta(r)$ curves are essentially the same. This is of course what one expects.
On the other hand one also expects some differences coming from the nonlinear regime of (\ref{fulleq}), and those are 
also present if we enlarge the very near horizon region of the individual curves. The main features of these
differences are plotted on Fig.~3 and Fig.~4.   
\begin{figure}[!ht]
\noindent\hfil\includegraphics[scale=1]{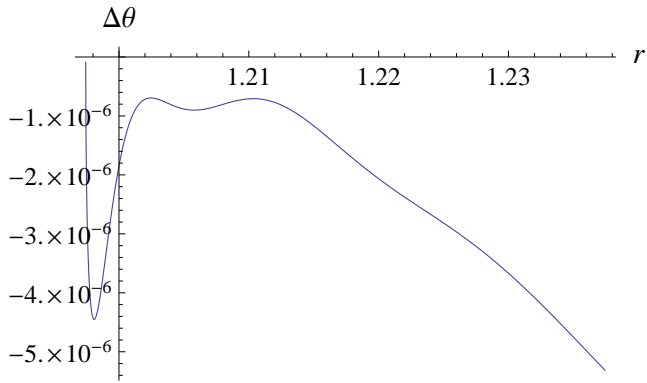} 
\caption{Near horizon nonlinear effects on a $\Delta\theta(r)$ curve in 4 dimensions with minimum 
horizon distance $r_1=1.19744$.}
\end{figure}
\begin{figure}[!ht]
\noindent\hfil\includegraphics[scale=1]{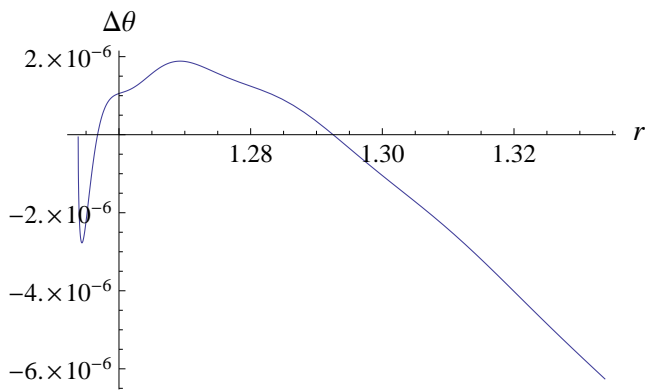} 
\caption{Near horizon nonlinear effects on a $\Delta\theta(r)$ curve in 4 dimensions with minimum 
horizon distance $r_1=1.2538$.}
\end{figure}

On Fig.~3 an interesting nonlinear effect appears compared to the perturbative solution. 
Instead of monotonically decreasing down to its global minimum, as $\vp$ does in the perturbative case,
the $\Delta\theta(r)$ curve has an extra local pattern near the minimum horizon distance $r_1$. 
The extra pattern is that the curve goes through some additional local extrema in this region before it 
finally tends to reach the global minimum which is similar to the one in the perturbative case. 
During these local differences however, $\Delta\theta(r)$ remains negative in this whole region.

On the qualitatively similar curve on Fig.~4, the essential difference compared to Fig.~3 is that 
$\Delta\theta(r)$ develops a sign
change in this region, that is there are several intersections in the very near horizon region between
the thick and thin solutions compared to the single intersection that is present in the case of the perturbative
solution (see Fig.~6 in \cite{CF}). 

With increasing horizon distance $r$, the above nonlinearities decay quickly, and
the solution agrees very well with the perturbative results.    

\section{Conclusions}

In the present work we studied the problem of a topologically flat 2-brane in a higher dimensional, thickness 
corrected BBH system. We provided a regular, non-perturbative, numerical solution for this special case based on 
earlier perturbative considerations \cite{CF}. The main result of this paper is the observation that the 2-dimensional case 
of the problem is a special one as being non-analytic at the axis of the system. This property makes it unique in 
the family of thick solutions, as in all other dimensions both the Minkowski and black hole embedding solutions 
are analytic in their entire domain. 

We analyzed the asymptotic behavior of the solution and obtained that it is at most a $C^1$, but in the general case 
it is only a $C^0$ function at the axis, just like the corresponding thin solutions. The initial conditions
of the problem are not uniquely fixed by regularity requirements, thus the solution we provided here is not unique. It
is however perfectly consistent with our earlier perturbative results.

With the present paper, we have provided the complete set of solutions of the thick-BBH problem in a series of
three consecutive papers. First, in \cite{CF}, we obtained all possible perturbative solutions and later, in 
\cite{Cz} all non-perturbative solutions were given except the case of the flat 2-brane. Present work completes 
the set.

In \cite{CF} we analyzed the properties of a topology changing, quasi-static phase transition in the thick-BBH
system. The obtained results in the present case however does not change our previous findings and thus we 
don't consider the phase transition in this paper.

The result, that thickness corrections change the analytic properties of the brane solutions at the axis of the 
system might have some physical consequences. Infinitely thin brane solutions, naturally, are very important in 
higher dimensional physics, but considering the present problem, one has the intuition that the thickness
corrections, which are in agreement with field theoretic domain wall models, made the 
thin-BBH system more stable in the sense that regular, analytic solutions could be provided in
essentially all cases. Since small physical perturbations to any system are usually proportional to the
derivatives of the unperturbed solution, it is very possible that the thin-BBH solutions are not entirely stable 
against small perturbations in the Minkowski branch. (Stability properties of the BBH system for the analytic
black hole embedding
solutions has been studied for example in \cite{HMN}.) This property however has been cured by the thickness corrections
and it is somehow in accordance with our physical expectations.

The special case of the 2-brane as ``remaining`` non-analytic after thickness corrections, thus, is an unexpected property,
which makes it physically interesting. So much the more that thick 2-branes, i.e.~thin walls on black hole backgrounds
in standard 4-dimensional general relativity are certainly real, physical objects. The fact that these solutions are 
essentially different from the corresponding ones in higher dimensions is remarkable.     
 
\acknowledgments
Most part of the calculations were performed and checked using the computer algebra program 
MATHEMATICA 7. The research was supported by the National Research Foundation of South Africa 
and the Hungarian National Research Fund, OTKA No.~K67790 grant.

\appendix*
\section{Coefficient Functions}
\begin{widetext}
{\small
\begin{eqnarray}
T_2&=&\frac{1}{64(a+b)\ve f^2r^4F^4}\left[\right.(\ve \dot f^3 r^6 \dot\theta^3 (-8 (10 a + 7 b) 
+ 18 (2 a + 3 b) f r^2 \dot\theta^2 \\
&+& 6 (2 a + b) f^2 r^4 \dot\theta^4+(a + b) f^3 r^6 \dot\theta^6) - 2 \ve \dot f^2 r^3 (-12 (10 a + 7 b) 
\dot\theta + 2 (126 a + 109 b) f r^2 \dot\theta^3 \nonumber\\
&+& 
90 a f^2 r^4 \dot\theta^5 + 3 (13 a + 5 b) f^3 r^6 \dot\theta^7 + 
2 (3 a + b) f^4 r^8 \dot\theta^9 - 
4 (18 a + 17 b) r \ddot\theta + (5 a + b) f^3 r^7 \dot\theta^8 \cot\theta \nonumber\\ &+& 
2 r \dot\theta^2 (3 (46 a + 43 b) f r^2 \ddot\theta + (-26 a - 7 b) \cot\theta) - 
f^2 r^5 \dot\theta^6 \csc\theta (-23 (a + b) \cos\theta + (5 a + b) f r^2 \ddot\theta \sin\theta) \nonumber\\
&-& 
2 f r^3 \dot\theta^4 \csc\theta ((17 a - 4 b) \cos\theta + 3 (7 a + 10 b) f r^2 \ddot\theta \sin\theta)) - 
4 \dot f r (f^4 r^9 (2 a \ve + 2 (8 a + 3 b) \ve f + r^2  \nonumber\\ 
&+& (a + b) \ve \ddot f r^2) \dot\theta^9 - 12 (2 a + b) \ve f^4 r^8 \dot\theta^8 \cot\theta - 
     4 \ve (3 (20 a + 19 b) f r^2 \ddot\theta + b \cot\theta) + 
     2 r \dot\theta (r^2 - 4 (2 a + b) \ve \ddot f r^2  \nonumber\\ 
&+& 
        180 (a + b) \ve f^2 r^4 \ddot\theta^2 + 7 a \ve \cot^2\theta - b \ve \cot^2\theta - 
        2 \ve f (6 (5 a + 4 b) + (28 a + 19 b) r^2 \ddot\theta \cot\theta) - 
        8 a \ve \csc^2\theta)  \nonumber\\ 
&+& 
     f^3 r^7 \dot\theta^7 (16 a \ve + 4 b \ve + 18 (5 a + 2 b) \ve f + 
        5 r^2 + (9 a + 5 b) \ve \ddot f r^2 + 5 a \ve \cot^2\theta + 
        b \ve \cot^2\theta - 4 a \ve \csc^2\theta)  \nonumber\\ 
&+& 
     f r^3 \dot\theta^3 (17 a \ve + 7 b \ve + 7 r^2 + 12 b \ve \ddot f r^2 - 
        60 (a + b) \ve f^2 r^4 \ddot\theta^2 + 38 a \ve \cot^2\theta + 
        2 \ve f (2 (82 a + 63 b) \nonumber\\ 
&-&  (47 a + 26 b) r^2 \ddot\theta \cot\theta) - 
        41 a \ve \csc^2\theta - 3 b \ve \csc^2\theta) + 
     2 \ve f^2 r^4 \dot\theta^4 \csc\theta ((-127 a - 55 b) \cos\theta \nonumber\\ 
&+&  (29 a - 7 b) f r^2 \ddot\theta \sin\theta) + 
     2 \ve f^3 r^6 \dot\theta^6 \csc\theta ((-49 a - 18 b) \cos\theta + 6 (2 a + b) f r^2 \ddot\theta \sin\theta)  \nonumber\\ 
&+& 
     6 \ve f r^2 \dot\theta^2 \csc\theta (-15 (2 a + b) \cos\theta + 2 (47 a + 43 b) f r^2 \ddot\theta \sin\theta) + 
     f^2 r^5 \dot\theta^5 \csc\theta ((2 a \ve + 8 b \ve + 9 r^2  \nonumber\\ 
&+& 24 (a + b) \ve \ddot f r^2) \sin\theta + 
        3 \ve f (2 (3 a + 4 b) r^2 \ddot\theta \cos\theta + 
           3 (23 a + 7 b) \sin\theta))) - 
  8 (-\cot\theta (r^2 - 2 a \ve \ddot f r^2 \nonumber\\ 
&+&  (a + b) \ve \cot^2\theta - 
        2 a \ve \csc^2\theta - 2 b \ve \csc^2\theta) - 
     f r (r \ddot\theta (-r^2 + 
           2 (10 a + 9 b) \ve \ddot f r^2 + (-7 a + b) \ve \cot^2\theta  \nonumber\\ 
&+& 
           8 a \ve \csc^2\theta) + 
        r \dot\theta^2 \cot\theta (3 a \ve - b \ve + 4 r^2 + (5 a + 4 b) \ve \ddot f r^2 + 
           4 (a + b) \ve \cot^2\theta - 9 a \ve \csc^2\theta \nonumber\\ 
&-&  9 b \ve \csc^2\theta) + 
        \dot\theta (12 (3 a + 2 b) \ve \ddot f r^2 + 2 (2 a + b) \ve f^{(3)} r^3 + 
           1/2 ((9 a + b) \ve - 
              3 r^2  \nonumber\\ 
&+& ((-7 a + b) \ve + 3 r^2) \cos(2\theta)) \csc^2\theta)) + 
     2 f^5 r^7 \dot\theta^6 \csc\theta (3 (2 a + b) \ve r \dot\theta^2 \cos\theta - 3 (2 a + b) \ve \dot\theta \sin\theta \nonumber\\ 
&+&  
        r^2 (2 a \ve + b \ve + r^2 + a \ve \ddot f r^2) \dot\theta^3 \sin\theta - 
        3 (2 a + b) \ve r \ddot\theta \sin\theta) - 
     f^2 r (r^2 (-2 a \ve - 6 b \ve - 11 r^2 + 2 (24 a + 13 b) \ve \ddot f r^2 \nonumber\\ 
&+& 
           2 (5 a + 2 b) \ve f^{(3)} r^3) \dot\theta^3 + 
        r \dot\theta^2 (r^2 (4 a \ve - 8 b \ve - 3 r^2 + 6 (a + b) \ve \ddot f r^2) \ddot\theta + 
           3 (15 a + 7 b) \ve \cot\theta) \nonumber\\
&+& 
        6 \ve \dot\theta (a + b + (15 a + 11 b) r^2 \ddot\theta \cot\theta) + 
        r^3 \dot\theta^4 \cot\theta (4 (5 a + 2 b) \ve \ddot f r^2 - 
           3/2 (5 a \ve + 9 b \ve - 
              2 r^2 \nonumber\\ 
&+& (a \ve - 3 b \ve + 2 r^2) \cos(2\theta)) \csc^2\theta) + 
        6 (a + b) \ve r \ddot\theta \csc\theta (2 r^2 \ddot\theta \cos\theta + 9 \sin\theta))  \nonumber\\ 
&-& 
     1/2 f^4 r^5 \dot\theta^2 (240 (a + b) \ve r^2 \dot\theta \ddot\theta^2 + 
        240 (a + b) \ve r^3 \ddot\theta^3 - 
        4 \ve \dot\theta^3 (-9 b + (20 a + 17 b) r^2 \ddot\theta \cot\theta)  \nonumber\\ 
&+& 
        2 r^2 \dot\theta^5 (-26 a \ve - 16 b \ve - 9 r^2 + 3 (-a + b) \ve \ddot f r^2 + 
           2 a \ve f^{(3)} r^3 - 2 a \ve \cot^2\theta - 8 b \ve \cot^2\theta + 
           6 b \ve \csc^2\theta)  \nonumber\\ 
&-& 
        2 r \dot\theta^4 (2 (22 a + 13 b) \ve \cot\theta + 
           r^2 \ddot\theta (8 a \ve + 8 b \ve + r^2 + 4 a \ve \ddot f r^2 - 
              2 (a - b) \ve \cot^2\theta + 4 a \ve \csc^2\theta)) \nonumber\\ 
&-&  
        12 \ve r \dot\theta^2 \ddot\theta \csc\theta (8 (a + b) r^2 \ddot\theta \cos\theta + (-8 a - 11 b) \sin\theta) + 
        r^3 \dot\theta^6 \cot\theta \csc^2\theta (-2 a \ve - 6 b \ve + 
           r^2  \nonumber\\ 
&-& (2 a \ve - 2 b \ve + r^2) \cos(2\theta) + 
           8 a \ve \ddot f r^2 \sin^2\theta)) - 
     1/2 f^3 r^3 (-600 (a + b) \ve r^2 \dot\theta \ddot\theta^2 - 
        40 (a + b) \ve r^3 \ddot\theta^3  \nonumber\\ 
&+& 
        2 \ve \dot\theta^3 (-3 (37 a + 29 b) + 2 (25 a + 16 b) r^2 \ddot\theta \cot\theta) - 
        2 r \dot\theta^4 ((-13 a - b) \ve \cot\theta + 
           r^2 \ddot\theta (16 a \ve + 16 b \ve + 3 r^2 \nonumber\\ 
&+&  6 (3 a + 2 b) \ve \ddot f r^2 + 
              3 (a + b) \ve \cot^2\theta)) + 
        2 r^2 \dot\theta^5 (-35 a \ve - 23 b \ve - 
           15 r^2 + (11 a + 5 b) \ve \ddot f r^2 + 2 (4 a + b) \ve f^{(3)} r^3  \nonumber\\ 
&-& 
           6 a \ve \cot^2\theta - 12 b \ve \cot^2\theta + 3 a \ve \csc^2\theta + 
           9 b \ve \csc^2\theta) - 
        2 \ve r \dot\theta^2 \ddot\theta \csc\theta (36 (a + b) r^2 \ddot\theta \cos\theta \nonumber\\ 
&+&  (295 a + 271 b) \sin\theta) - 
        r^3 \dot\theta^6 \cot\theta \csc^2\theta (9 a \ve + 21 b \ve - 
           4 r^2 + (5 a \ve - 7 b \ve + 4 r^2) \cos(2\theta)  \nonumber\\ 
&-& 
           2 (17 a + 4 b) \ve \ddot f r^2 \sin^2\theta)))) \left.\right],\nonumber
\end{eqnarray}
\begin{eqnarray}
A_4&=&\frac{1}{188697600A_1^5(a+b)\ve r^6}\left[\right.
32 A_1^3 r^3 (-15263640 A_2^3 (a + b) \ve r^3 + 
    1260 A_1^2 \ve r (-491 a A_2 - 608 A_2 b\\
&+& 2573 a A_3 r + 
       2540 A_3 b r) - 
    5040 A_1 A_2 \ve r^2 (A_2 (497 a + 503 b) - 4606 A_3 (a + b) r)  \nonumber\\ 
&+&
    420 A_1^5 r (-65 a \ve - 38 b \ve + 6 r^2) + 
    A_1^9 r^3 (-6 a \ve - 34 b \ve + 7 r^2) + 
    14 A_1^7 r^2 (46 a \ve + 40 b \ve + 15 r^2) \nonumber\\ 
&+&
    21 A_1^6 A_2 r^3 (222 a \ve - 38 b \ve + 65 r^2) - 
    630 A_1^3 (2 (71 a + 176 b) \ve + 
       A_2^2 r^3 (27 a \ve - 41 b \ve + 17 r^2)) \nonumber\\ 
&+&
    210 A_1^4 r^2 (A_3 r (983 a \ve + 1067 b \ve - 21 r^2) + 
       A_2 (490 a \ve + 160 b \ve + 93 r^2))) \nonumber\\ 
&+& (1/(f^3))
 7 (-33177600 (a + b) \ve + 
    A_1 r (90 A_1 \ve \dot f r (-512 (616 a + 601 b) - 
          128 A_1^2 (611 a + 581 b) \dot f r^2  \nonumber\\ 
&-&
          32 A_1^4 (67 a + 119 b) \dot f^2 r^4 + 
          40 A_1^6 (3 a + b) \dot f^3 r^6 - A_1^8 (a + b) \dot f^4 r^8) + 
       15 f (-1536 (4 A_1 (646 a + 631 b) \ve \nonumber\\ 
&+& 7383 A_2 (a + b) \ve r + 
             A_1^3 r (19 (a + b) \ve + r^2)) + 
          A_1^2 r^2 (768 A_1 (267 a + 248 b) \ve \ddot f r  \nonumber\\ 
&+&
             6 A_1^6 (a + b) \ve \dot f^4 r^6 (4 A_1 + 15 A_2 r) + 
             A_1^4 \ve \dot f^3 r^4 (-48 A_1 (47 a + 17 b) + 
                8 (5 a (A_1^3 - 96 A_2) \nonumber\\ 
&+& (A_1^3 - 150 A_2) b) r + 
                21 A_1^3 (a + b) \ddot f r^3) + 
             16 \dot f (120 A_1^3 r^3 + 3 A_1^3 (491 a + 685 b) \ve \ddot f r^3  \nonumber\\ 
&-&
                8 \ve (9138 a A_1 + 8802 A_1 b + 253 a A_1^3 r + 
                   28668 a A_2 r + 163 A_1^3 b r + 28038 A_2 b r))  \nonumber\\ 
&+&
             8 A_1^2 \dot f^2 r^2 (240 A_1 (9 a - 5 b) \ve - 
                48 A_2 (137 a + 148 b) \ve r - 
                3 A_1^3 (83 a + 29 b) \ve \ddot f r^3  \nonumber\\ 
&-&
                4 A_1^3 r (53 a \ve + 55 b \ve + 9 r^2)))) + 
       4 A_1^5 f^3 r^4 (90 A_1^2 \ve \dot f^2 r^2 (A_1^2 (41 a + 15 b) + 
             15 A_1 A_2 (25 a + 9 b) r  \nonumber\\ 
&+&
             5 (18 A_2^2 + 5 A_1 A_3) (3 a + b) r^2) + 
          3 \dot f (16 A_1^6 (2 a + b) \ve r^2 - 
             24480 A_2^2 (2 a + b) \ve r^2  \nonumber\\ 
&+&
             720 A_1 \ve r (87 a A_2 + 36 A_2 b - 14 A_3 (2 a + b) r) + 
             60 A_1^4 r (5 a \ve + 2 b \ve + 9 r^2) + 
             30 A_1^3 r^2 (25 A_3 r (2 a \ve + r^2)  \nonumber\\ 
&+&
                4 A_2 (97 a \ve + 41 b \ve + 30 r^2)) + 
             180 A_1^2 (-8 (2 a + b) \ve + 
                15 A_2^2 (2 a \ve r^3 + r^5)) + 
             15 A_1^2 \ve r^3 (2 \ddot f (3 A_1^2 (20 a + 7 b)  \nonumber\\ 
&+&
                   30 A_1 A_2 (7 a + 3 b) r + 
                   5 (18 A_2^2 + 5 A_1 A_3) (a + b) r^2) + 
                A_1 r (A_1 (a + b) f^{(4)} r + 
                   2 f^{(3)} (A_1 (11 a + 5 b)  \nonumber\\ 
&+& 15 A_2 (a + b) r)))) + 
          r (90 A_1^3 \ve \ddot f^2 r^3 (A_1 (9 a + 5 b) + 15 A_2 (a + b) r) + 
             \ddot f (135 A_1^4 (a + b) \ve f^{(3)} r^4 \nonumber\\ 
&+&
                270 A_1^3 r^3 (2 A_1 + 5 A_2 r) + 
                4 \ve (1080 A_1^2 (3 a + b) + 
                   15 A_1 (52 a A_1^3 - 1788 a A_2 + 15 A_1^3 b - 
                    828 A_2 b) r \nonumber\\ 
&+&
                   a (8 A_1^6 + 2235 A_1^3 A_2 - 12240 A_2^2 - 
                    5040 A_1 A_3) r^2)) + 
             15 A_1 r (-192 a A_1 \ve f^{(4)} r  \nonumber\\ 
&+&
                f^{(3)} (-48 A_1 (19 a + 6 b) \ve - 2208 a A_2 \ve r + 
                   A_1^3 (38 a \ve r + 3 r^3))))) - 
       A_1 f^2 r (384 (270 A_1^2 (249 a + 239 b) \ve  \nonumber\\ 
&+&
             619200 A_2^2 (a + b) \ve r^2 + 
             30 A_1 \ve r (15 A_2 (697 a + 683 b) - 6682 A_3 (a + b) r) + 
             30 A_1^4 r (77 a \ve + 55 b \ve - 3 r^2)  \nonumber\\ 
&+&
             15 A_1^3 A_2 r^2 (247 a \ve + 203 b \ve + 7 r^2) - 
             A_1^6 r^2 (23 a \ve + 3 b \ve + 10 r^2)) + 
          A_1^2 r^2 (90 A_1^4 (a + b) \ve \dot f^3 r^4 (6 A_1^2 + 
                90 A_2^2 r^2 \nonumber\\ 
&+& 5 A_1 r (12 A_2 + 5 A_3 r)) + 
             A_1^2 \ve \dot f^2 r^2 (8 (-360 A_1^2 (8 a + 3 b) + 
                   30 A_1 (5 a A_1^3 - 939 a A_2 + A_1^3 b - 
                    339 A_2 b) r \nonumber\\ 
&+& (A_1^6 + 195 A_1^3 A_2 - 1530 A_2^2 - 
                    630 A_1 A_3) (5 a + b) r^2) + 
                135 A_1^3 (a + b) r^3 (A_1 f^{(3)} r + 
                   \ddot f (8 A_1 + 30 A_2 r)))  \nonumber\\ 
&+&
             240 A_1 r (15 A_1 (49 a + 39 b) \ve f^{(3)} r + 
                \ddot f (24 A_1 (52 a - 15 b) \ve - 
                   6 A_2 (457 a + 587 b) \ve r - 
                   3 A_1^3 (19 a + 7 b) \ve \ddot f r^3  \nonumber\\ 
&-&
                   2 A_1^3 r (76 a \ve + 50 b \ve + 9 r^2))) + 
             6 \dot f (-720 A_1^3 r^3 (2 A_1 + 9 A_2 r) - 
                16 \ve (60 A_1^2 (105 a - 4 b) + 
                   20 A_1 (25 a A_1^3 \nonumber\\ 
&-& 330 a A_2 + 6 A_1^3 b - 
                    261 A_2 b) r + (a (7 A_1^6 + 1080 A_1^3 A_2 - 
                    30060 A_2^2 + 38265 A_1 A_3) + 
                    3 (3 A_1^6 + 190 A_1^3 A_2 \nonumber\\ 
&-& 9780 A_2^2 + 
                    13085 A_1 A_3) b) r^2) + 
                5 A_1^3 \ve r^3 (-24 A_1 (19 a + 7 b) f^{(3)} r + 
                   \ddot f (-48 A_1 (51 a + 19 b) \nonumber\\ 
&+&
                    8 (5 a A_1^3 - 588 a A_2 + A_1^3 b - 204 A_2 b) r + 
                    9 A_1^3 (a + b) \ddot f r^3))))) - 
       24 A_1^5 f^4 r^3 (4 (2 a + b) \ve (-180 A_1^2 \nonumber\\ 
&+&
             1620 A_1 A_2 r + (2 A_1^6 + 165 A_1^3 A_2 - 3060 A_2^2 - 
                1260 A_1 A_3) r^2 - 75 A_1^2 (18 A_2^2 + 5 A_1 A_3) r^3)  \nonumber\\ 
&-&
          60 A_1^2 r^3 (3 A_1^2 + 90 A_2^2 r^2 + 
             5 A_1 r (9 A_2 + 5 A_3 r)) - 
          15 A_1^2 \ve r^2 (2 (8 a + 3 b) \dot f (A_1^2 + 90 A_2^2 r^2 \nonumber\\ 
&+&
                5 A_1 r (6 A_2 + 5 A_3 r)) + 
             r (2 \ddot f (2 A_1^2 (11 a + 3 b) + 15 A_1 A_2 (14 a + 3 b) r \nonumber\\ 
&+&
                   10 a (18 A_2^2 + 5 A_1 A_3) r^2) + 
                A_1 r (2 a A_1 f^{(4)} r + 
                   f^{(3)} (3 A_1 b + 20 a (A_1 + 3 A_2 r))))))))
\left.\right]\left.\right|_{r_1}\ , \nonumber\\\nonumber\\\nonumber\\
A_2&=&\frac{1}{26880A_1(a+b)\ve fr^2}\left[\right.
(-14592 (a + b) \ve + 
  A_1^2 r (3 \ve \dot f r (-16 (29 a + 35 b) + 8 A_1^2 (5 a + b) \dot f r^2\\ 
&-&
        A_1^4 (a + b) \dot f^2 r^4)  
     24 A_1^2 f^2 r (2 A_1^2 r^3 + 2 (2 a + b) \ve (12 + A_1^2 r) + 
        A_1^2 \ve r^2 ((8 a + 3 b) \dot f + 2 a \ddot f r)) \nonumber\\ 
&+&
     4 f (-8 (6 (31 a + 28 b) \ve + 
           A_1^2 r (19 a \ve + 7 b \ve + 3 r^2)) + 
        3 A_1^2 r^2 (\ve \ddot f r (-32 a + A_1^2 (a + b) \dot f r^2) \nonumber\\ 
&+&
           \dot f (-48 (2 a + b) \ve + A_1^2 (3 a + b) \ve \dot f r^2 + 
              A_1^2 r (2 a \ve + r^2))))))
\left.\right]\left.\right|_{r_1}\ , \nonumber
\end{eqnarray}
\begin{eqnarray}
A_3&=&\frac{1}{1140480A_1^3(a+b)\ve r^4}\left[\right.
(15 \ve (56576 (a + b) + 384 A_1^2 (74 a + 69 b) \dot f r^2 + 
     32 A_1^4 (37 a + 57 b) \dot f^2 r^4\\
&-& 8 A_1^6 (10 a + 3 b) \dot f^3 r^6 + 
     A_1^8 (a + b) \dot f^4 r^8))/f^2 + 
  120 A_1^5 f r^3 (6 A_1^2 r^3 (A_1 + 5 A_2 r)  \nonumber\\ 
&-&
     2 (2 a + b) \ve (12 A_1 + A_1^3 r - 84 A_2 r - 15 A_1^2 A_2 r^2) + 
     A_1^2 \ve r^2 ((8 a + 3 b) \dot f (2 A_1 + 15 A_2 r) \nonumber\\ 
&+&
        r (2 a A_1 f^{(3)} r + \ddot f (14 a A_1 + 3 A_1 b + 30 a A_2 r)))) + (1/f)
  5 A_1 r (64 (30 A_1 (101 a + 95 b) \ve  \nonumber\\ 
&+& 9372 A_2 (a + b) \ve r + 
        A_1^3 r (79 a \ve + 67 b \ve - 9 r^2)) + 
     A_1^2 r^2 (-144 A_1 (35 a + 33 b) \ve \ddot f r  \nonumber\\ 
&-&
        3 A_1^4 (a + b) \ve \dot f^3 r^4 (4 A_1 + 15 A_2 r) + 
        A_1^2 \ve \dot f^2 r^2 (24 A_1 (37 a + 13 b)  -
           4 (A_1^3 - 42 A_2) (5 a + b) r \nonumber\\ 
&-& 9 A_1^3 (a + b) \ddot f r^3) + 
        16 \dot f (12 A_1 (-30 a + 13 b) \ve + 3 A_2 (517 a + 523 b) \ve r + 
           12 A_1^3 (3 a + b) \ve \ddot f r^3  \nonumber\\ 
&+&
           A_1^3 r (40 a \ve + 34 b \ve + 9 r^2)))) + 
  4 A_1^2 r^2 (4 (180 A_1^2 (25 a + 32 b) \ve + 
        180 A_1 A_2 (173 a + 172 b) \ve r  \nonumber\\ 
&+& 123480 A_2^2 (a + b) \ve r^2 + 
        30 A_1^3 A_2 r^2 (37 a \ve + 65 b \ve - 7 r^2) + 
        30 A_1^4 r (30 a \ve + 16 b \ve + 3 r^2) \nonumber\\ 
&+&
        A_1^6 r^2 (6 a \ve - 14 b \ve + 5 r^2)) + 
     5 A_1^3 r^2 (3 A_1^2 \ve \dot f^2 r^2 (A_1 (25 a + 9 b) + 
           15 A_2 (3 a + b) r) \nonumber\\ 
&+&
        3 \dot f (8 A_1 (29 a + 12 b) \ve - 336 A_2 (2 a + b) \ve r + 
           15 A_1^2 A_2 r^2 (2 a \ve + r^2) + 
           4 A_1^3 r (7 a \ve + 3 b \ve + 2 r^2)  \nonumber\\ 
&+&
           A_1^2 \ve r^3 (A_1 (a + b) f^{(3)} r + 
              \ddot f (2 A_1 (7 a + 3 b) + 15 A_2 (a + b) r))) + 
        r (-144 a A_1 \ve f^{(3)} r \nonumber\\ 
&+&
           \ddot f (-72 A_1 (7 a + 3 b) \ve - 672 a A_2 \ve r + 
              3 A_1^3 (a + b) \ve \ddot f r^3 + A_1^3 (22 a \ve r + 3 r^3)))))
\left.\right]\left.\right|_{r_1}\ . \nonumber
\end{eqnarray}}
\end{widetext}

\end{document}